\documentclass[a4paper,12pt]{article}
\usepackage{epsfig}
\usepackage{citesort}
\usepackage{psfrag}
\date{\today}
 \normalsize

\newcommand{\bmat}{\left(\begin{array}}
\newcommand{\emat}{\end{array}\right)}
\newcommand{\be}{\begin{equation}}
\newcommand{\ee}{\end{equation}}
\newcommand{\bea}{\begin{eqnarray}}
\newcommand{\eea}{\end{eqnarray}}

\def\lsim{\raise0.3ex\hbox{$\;<$\kern-0.75em\raise-1.1ex\hbox{$\sim\;$}}}
\def\gsim{\raise0.3ex\hbox{$\;>$\kern-0.75em\raise-1.1ex\hbox{$\sim\;$}}}
\def\Frac#1#2{\frac{\displaystyle{#1}}{\displaystyle{#2}}}

\def\ep{\eta^{\prime}}
\def\lsim{\raise0.3ex\hbox{$\;<$\kern-0.75em\raise-1.1ex\hbox{$\sim\;$}}}
\def\gsim{\raise0.3ex\hbox{$\;>$\kern-0.75em\raise-1.1ex\hbox{$\sim\;$}}}
\def\Frac#1#2{\frac{\displaystyle{#1}}{\displaystyle{#2}}}

\def\susy{\mbox{\tiny SUSY}}
\def\sm{\mbox{\tiny SM}}
\def\du#1#2{{\left(\delta^u_{#1}\right)_{#2}}}

\newcommand{\W}{{\scriptscriptstyle W}}

\def\dofigs#1#2#3{\centerline{\epsfxsize=#1\epsfig{file=#2, width=7.5cm,
height=7.5cm, angle=-90}
\hfil\epsfxsize=#1\epsfig{file=#3,  width=7.5cm, height=7.5cm, angle=-90}}}

\def\dofigs#1#2#3{\centerline{\epsfxsize=#1\epsfig{file=#2, width=6cm,
height=7.5cm, angle=-90}\hspace{0cm}
\hfil\epsfxsize=#1\epsfig{file=#3,  width=6cm, height=7.5cm, angle=-90}}}
\begin{document}
\renewcommand{\thefootnote}{\fnsymbol{footnote}}
{\large
\begin{center}
{\bf Supersymmetric contributions to the CP asymmetry of
the $B \to \phi K_S$ and $B\to \eta' K_S$ }

\end{center}}
\vspace{.3cm}

\begin{center}
Shaaban Khalil\\
\vspace{.3cm} $^1$ \emph{IPPP, University of Durham, South Rd.,
Durham DH1 3LE, U.K.}\\
$^2$ \emph{Department of Mathematics, German University in Cairo,
New Cairo city,
El Tagamoa El Khames, Egypt.}\\
\end{center}

\vspace{.3cm}
\hrule \vskip 0.3cm
\begin{center}
\small{\bf Abstract}\\[3mm]
\end{center}
We analyse the CP asymmetry of the $B \to \phi K_S$ and $B\to \eta' K_S$
processes in general supersymmetric models. We consider both  gluino and chargino
exchanges in a model independent way by using the mass insertion
approximation method. We adopt the QCD factorization method for evaluating the
corresponding hadronic matrix elements. We show that chromomagnetic type
of operator may play an important role in accounting for the deviation of the
mixing CP asymmetry between $B \to \phi K_S$ and $B \to J/\psi K_S$ processes
observed by Belle and BaBar experiments.
We also show that due to the different parity in the final states of these processes,
their supersymmetric contributions from the R-sector have an opposite sign, which naturally
explain the large deviation between their asymmetries.
\begin{minipage}[h]{14.0cm}
\end{minipage}
\vskip 0.3cm \hrule \vskip 0.5cm
%
\section{Introduction}
One of the most important tasks for B factory experiments would be to test
the Kobayashi-Maskawa (KM) ansatz for the flavor CP violation. The flavor CP violation
has been studied quite a while, however, it is still one of the least tested aspect
in the standard model (SM). Although it is unlikely that the SM provides the
complete description of CP violation in nature (e.g. Baryon asymmetry in the universe),
it is also very difficult to
include any additional sources of CP violation beyond the phase in the CKM mixing
matrix. Stringent constraints on these phases are usually obtained from the
experimental bounds on the electric dipole moment (EDM) of the neutron, electron
and mercury atom. Therefore, it remains a challenge for any new physics beyond the
SM to give a new source of CP violation that may explain possible deviations
from the SM results and also avoid overproduction of the EDMs. In supersymmetric
theories, it has been emphasised ~\cite{Abel:2001vy} that there are attractive scenarios
where the EDM problem is solved and genuine SUSY CP violating effects are found.

Recently, BaBar and Belle collaborations \cite{Bfact} announced
large deviations from the SM expectations in the CP asymmetry of
$B^0\to \phi K_S$ and branching ratio of $B^0\to \eta^{\prime} K^0$.
These discrepancies have been interpreted as possible consequences
of new physics (NP) beyond the SM
\cite{PhiKs_NP,KM,PhiKs_Rparity,PhiKs_gluino,PhiKs_sugra,CFMS,KK,chargino,emidio1,emidio2,hpenguin}.
For $B^0$ and $\bar{B}^0$ decays to a CP eigenstate $f_{CP}$, the
time dependent CP asymmetries are usually described by rates
$a_{f_{CP}}(t)$,
\begin{eqnarray}
a_{f_{CP}}(t)=\frac{\Gamma (\overline{B}^0(t)\to f_{CP})-\Gamma
(B^0(t)\to f_{CP})} {\Gamma (\overline{B}^0(t)\to f_{CP})+\Gamma
(B^0(t) \to f_{CP})} =C_{f_{CP}}\cos\Delta
M_{B_d}t+S_{f_{CP}}\sin\Delta M_{B_d}t, \label{CPasym}
\end{eqnarray}
where $C_{f_{CP}}$ and $S_{f_{CP}}$ represent the parameters of
direct  and indirect CP violations respectively, and $\Delta
M_{B_d}$ is the $B^0$ eigenstate mass difference.

In the SM, the decay process of $B \to \phi K$ is dominated by the
top quark intermediated penguin diagram, which do not include any CP
violating phase. Therefore, the CP asymmetry of $B \to J/\psi K_S$
and $B \to \phi K_S$ in SM are caused only by the phase in
$B^0-\overline{B}^0$ mixing diagram and we expect $S_{J/\psi K_S} =
S_{\phi K_S}$ where $S_{f_{CP}}$ represents the mixing CP asymmetry.
The $B \to \ep K_S$ process is induced by more diagrams since $\ep$
meson contains not only $s\bar{s}$ state but also $u\bar{u}$ and
$d\bar{d}$ states with the pseudoscalar mixing angle $\theta_p$.
Nevertheless, under an assumption that its tree diagram contribution
is very small, which is indeed the case, one can expect $S_{\phi
K_S} =S_{\ep K_S}$ as well. Thus, the series of new experimental
data surprised us: \bea S_{J/\psi K_S} =0.726\pm0.037, \label{Spsi}
\eea which agrees quite well with the SM prediction
$0.715_{-0.045}^{+0.055}$ \cite{AB}. However, results of Belle on
the corresponding $\sin{2\beta}$ extracted for $B^0\to \phi K_S$
process has changed dramatically \cite{giorgi,sakai} \bea S_{\phi
K_S}&=&0.50\pm 0.25^{+0.07}_{-0.04}\;\;
({\rm BaBar}),\nonumber\\
&=&0.06\pm 0.33 \pm 0.09\;\; ({\rm Belle})\, , \label{Sphi} \eea
where the first errors are statistical and the second systematic,
showing now a better agreement than before
\cite{phi_babar,phieta_belle}. However, as we can see from
Eq.(\ref{Sphi}), the relative central values are still different.
BaBar results \cite{giorgi} are more compatible with SM predictions,
while Belle measurements \cite{sakai} still show a deviation from
the $c\bar{c}$ measurements of about $2\sigma$. Moreover, the
average $S_{\phi K_S}=0.34 \pm 0.20$ is quite different from the
previous one \cite{hfag}, displaying now 1.7$\sigma$ deviation from
Eq.(\ref{Spsi}).

Furthermore, the most recent measured CP asymmetry in the $B^0\to
\eta^{\prime} K_S$ decay is found by BaBar \cite{giorgi} and Belle
\cite{sakai} collaborations as \bea
S_{\eta^{\prime} K_S}&=&0.27\pm 0.14\pm 0.03 \;\; ({\rm BaBar}) \nonumber\\
&=&0.65\pm 0.18\pm 0.04 \;\; ({\rm Belle}), \label{Seta} \eea with
an average $S_{\eta^{\prime} K_S}= 0.41 \pm 0.11$, which shows a
2.5$\sigma$  discrepancy from Eq.~(\ref{Spsi}). For the previous
results see (BaBar) \cite{etaBa} and (Belle) \cite{phieta_belle}.

It is interesting to note that the new results on s-penguin modes
from both experiments differ from the value extracted from the
$c\bar{c}$ mode ($J/\psi$), BaBar by 2.7$\sigma$ and Belle by
2.4$\sigma$ \cite{giorgi,sakai}. At the same time the experiments
agree with each other, and even the central values are quite close:
\bea 0.42\pm0.10\;\; {\rm BaBar},\;\;\; 0.43^{+0.12}_{-0.11}\;\;{\rm
  Belle}.
\nonumber \eea

Supersymmetry (SUSY) is one of the most popular candidates for
physics beyond the SM. In SUSY models there are new sources of CP
violation beside the CKM phase \cite{GNR}. In this review we show
our attempts to understand all the above experimental data within
the Supersymmetric models.

\section{The mass insertion approximation}
As mentioned, the SUSY extension of the SM may provide considerable effects to
the CP violation observables since it contains new CP violating phases and also new
flavour structures. Thus, SUSY is a natural candidate to resolve the discrepancy among the
observed mixing CP asymmetries in $B$-meson decays.

In the following, we will perform a model independent
analysis by using the mass insertion approximation \cite{HallRaby}. We
start with the minimal supersymmetric standard model (MSSM),
where a minimal number of super-fields is introduced and $R$ parity
is conserved, with the following soft SUSY breaking terms
\bea\label{susy:gen:vsb}
V_{SB} &=& m_{0\alpha}^2 \phi_{\alpha}^* \phi_{\alpha} +
\epsilon_{ab}
\Big(A^u_{ij} Y^u_{ij} H_2^b \tilde{q}_{L_i}^a \tilde{u}^*_{R_j} +
A^d_{ij} Y^d_{ij} H_1^a \tilde{q}_{L_i}^b \tilde{d}^*_{R_j} \nonumber\\
&+& A^l_{ij} Y^l_{ij} H_1^a \tilde{l}_{L_i}^b \tilde{e}^*_{R_j} - B\mu H_1^a H_2^b +
\mathrm{H.c.} \Big)\nonumber\\
&-& \frac{1}{2} \Big(m_3\bar{\tilde{g}} \tilde{g} +
m_2 \overline{\widetilde{W^a}} \widetilde{W}^a +
m_1 \bar{\tilde{B}} \tilde{B}\Big)\;,
\eea
where $i,j$ are family indices, $a,b$ are $SU(2)$ indices, and
$\epsilon_{ab}$ is the $2\times 2$ fully antisymmetric tensor, with
$\epsilon_{12}=1$. Moreover, $\phi_{\alpha}$ denotes all
the scalar fields of the theory.
Although in general the parameters $\mu$,
$B$, $A^\alpha$ and $m_i$ can be complex, two of their
phases can be rotated away.

The mass insertion approximation is a technique which is developed to include the
soft SUSY breaking term without specifying the models in behind. In this approximation,
one adopts a basis where the couplings of the fermion and sfermion are flavour diagonal, leaving
all the sources of flavour violation inside the off-diagonal terms of the sfermion mass
matrix. These terms are denoted by $(\Delta^{q}_{AB})^{ij}$,
where $A,B=(L,R)$ and $q=u,d$. The sfermion propagator
is then expanded as
\begin{equation}
\langle \tilde q_A^a \tilde q_B^{b*} \rangle ={\rm i}~(k^2 {\bf 1}-\tilde{m}^2 {\bf 1}-
\Delta^q_{AB})^{-1}_{a b}\simeq
{{\rm i}~\delta_{ab}\over k^2-\tilde{m}^2}  +{{\rm i}~(\Delta^q_{AB})_{ab}\over
(k^2-\tilde{m}^2)^2},
\end{equation}
where ${\bf 1}$ is the unit matrix and $\tilde{m}$ is the average squark mass.
The SUSY contributions
are parameterised in terms of the dimensionless parameters
 $(\delta^{q}_{A B})_{ij}=
(\Delta^{q}_{AB})^{ij}/\tilde{m}^2$.
This method allows to parametrise, in a
model independent way, the main sources of flavor violations in SUSY models.

Including the SUSY contribution, the effective Hamiltonian $H^{\Delta B=1}_{\rm eff}$
for these processes can be expressed via the Operator
Product Expansion (OPE) as
\bea
H^{\Delta B=1}_{\rm eff}&=&\left\{ \frac{G_F}{\sqrt{2}}
\sum_{p=u,c} \lambda_p ~\left( C_1 Q_1^p + C_2 Q_2^p +
\sum_{i=3}^{10} C_i Q_i + C_{7\gamma}
Q_{7\gamma} + C_{8g} Q_{8g} \right) + h.c. \right\}
\nonumber\\
&+&\left\{Q_i\to \tilde{Q}_i\, ,\, C_i\to \tilde{C}_i\right\}
\;,
\label{Heff}
\eea
where $\lambda_p= V_{pb} V^{\star}_{p s}$,
with $V_{pb}$ the unitary CKM matrix elements satisfying
$\lambda_t+\lambda_u+\lambda_c=0$, and $C_i\equiv C_i(\mu_b)$ are
the Wilson coefficients at low energy scale
$\mu_b\simeq m_b$.

As emphasised in \cite{KK,emidio1}, the leading contribution of both gluino and chargino to
$\Delta B=1$ processes come from the chromomagnetic penguin operator $O_{g} (\tilde{O}_g)$.
The corresponding Wilson coefficient is given by
\begin{eqnarray}
C^{\tilde{g}}_{8g} \!\! &=&\!\!\frac{ \alpha_s \pi}
{\sqrt{2}G_F m_{\tilde{q}}^2}\!\left[
\!(\delta_{LL}^d)_{23}\left( \frac{1}{3} M_3(x)\! + \!3 M_4(x)\right)\!+\!
(\delta_{LR}^d)_{23}\frac{m_{\tilde{g}}}{m_b}
\left(\!\frac{1}{3} M_1(x) \!+\! 3 M_3(x)\right)\!\right]\!,
\end{eqnarray}
and
\begin{eqnarray}
C_{8g}^{\chi}=\big[\du{LL}{32}+\lambda\du{LL}{31}\big] R_{8g}^{LL}
+\big[\du{RL}{32}+\lambda\du{RL}{31}\big]Y_t R_{8g}^{RL}.
\end{eqnarray}
Here the functions $R_{8g}^{LL}$ and $R_{8g}^{RL}$ are given by
\begin{eqnarray}
R_{8g}^{LL}&=&\sum_i |V_{i1}|^2\,
x_{Wi}\, P_{M^{\gamma,g}}^{LL}(x_i) - Y_b\sum_i V_{i1} U_{i2}\,
x_{Wi}\, \frac{m_{\chi_i}}{m_b} P_{M_{\gamma,g}}^{LR}(x_i),
\nonumber
\\
R_{8g}^{RL}&=& -\sum_i V_{i1}V_{i2}^{\star}\,
x_{Wi}\, P_{M_{\gamma,g}}^{LL}(x_i),
\label{Rterms}
\end{eqnarray}
where $x_{\W i}=m_W^2/m_{\chi_i}^2$,
$x_{i}=m_{\chi_i}^2/\tilde{m}^2$, $\bar x_i =\tilde{m}^2/m_{\chi_i}^2$, and
$x_{ij}=m_{\chi_i}^2/m_{\chi_j}^2$.
The loop functions $P_{8g}^{LL(LR)}(x)$ and also the functions $M_i(x)$, $i=1,3,4$ can be found
in Ref.\cite{emidio2}. Finally, $U$ and $V$ are the matrices that diagonalize chargino mass matrix.

It is now clear that the part proportional to LR mass insertions in $C^{\tilde{g}}_{8g}$
which is enhanced by a factor $m_{\tilde{g}}/m_b$ would give a dominant contribution. Also the part
proportional to the LL mass insertion in $C^{\chi}_{8g}$ is enhanced by $m_{\chi}/m_b$ and could also
give significant contribution.

\section{Can we explain the experimental data of $S_{\phi K_S}$ in SUSY? }
Following the parametrisation of the SM and SUSY amplitudes in Ref.\cite{KK},
$S_{\phi K_S}$ can be written as
\begin{eqnarray}
S_{\phi K_S}=\Frac{\sin 2 \beta +2 R_{\phi}
\cos \delta_{12} \sin(\theta_{\phi} + 2 \beta) +
R_{\phi}^2 \sin (2 \theta_{\phi} + 2 \beta)}{1+ 2 R_{\phi}
\cos \delta_{12} \cos\theta_{\phi} +R_{\phi}^2}
\end{eqnarray}
where $ R_{\phi}= \vert A^{\susy}/A^{\sm}\vert$, $\theta_{\phi}=
\mathrm{arg}(A^{\susy}/A^{\sm})$, and $\delta_{12}$ is the strong phase.

We will discuss in the following whether the SUSY contributions can
make $S_{\phi K_S}$ negative. For $m_{\tilde{q}}=m_{\tilde{g}}=500$\ GeV and adopting the
QCD factorization mechanism to evaluate the matrix elements, one obtains

\begin{equation}
R^{QCDF}_{\phi}|_{\tilde{g}} \simeq  \left\{-0.14\,\times e^{-i\,0.1}
 (\delta_{LR}^d)_{23}\,-\, 127\, \times e^{-i\,0.08}
(\delta_{LR}^d)_{23}\right\}
\,+\, \left\{L\leftrightarrow R\right\}
\label{Rgl_Phi},
\end{equation}
while in the case of chargino exchange with  gaugino mass $M_2=200$ GeV,
$\mu= 300$ GeV, and $\tilde{m}_{\tilde{t}_R}=150$ GeV,
we obtain, for $\tan{\beta}=40$
\begin{eqnarray}
R^{QCDF}_{\Phi}|_{\chi^ {\pm}} &\simeq&  1.89\times e^{-i\,0.07}
\, (\delta_{LL}^u)_{32}\,-\, 0.11\times e^{-i\,0.17}
\, (\delta_{RL}^u)_{32} \nonumber \\
&+&
0.43\times e^{-i\,0.07}
\, (\delta_{LL}^u)_{31}\,-\,
0.02\times e^{-i\,0.17}
\, (\delta_{RL}^u)_{31}.
\label{Rch_Phi}
\end{eqnarray}
From results in Eqs.(\ref{Rgl_Phi})--(\ref{Rch_Phi}),
it is clear that the largest SUSY
effect is provided by the gluino and chargino
contributions to the chromomagnetic operator which are proportional to
$(\delta_{LR}^d)_{23}$ and $(\delta_{LL}^u)_{32}$ respectively.
However, the $b\to s \gamma$ constraints play a crucial role in this case.
For the above SUSY configurations, the $b\to s \gamma$ decay set the following
constraints on gluino and chargino contributions, respectively
$\vert(\delta_{LR}^d)_{23}\vert< 0.016$ and $\vert(\delta_{LL}^u)_{32}\vert < 0.1$.
Implementing these bounds in Eqs.(\ref{Rgl_Phi})--(\ref{Rch_Phi}),
we see that gluino contribution can achieve larger value for $R_{\phi}$ than chargino one.

\begin{figure}[tpb]
\begin{center}
\dofigs{3.1in}{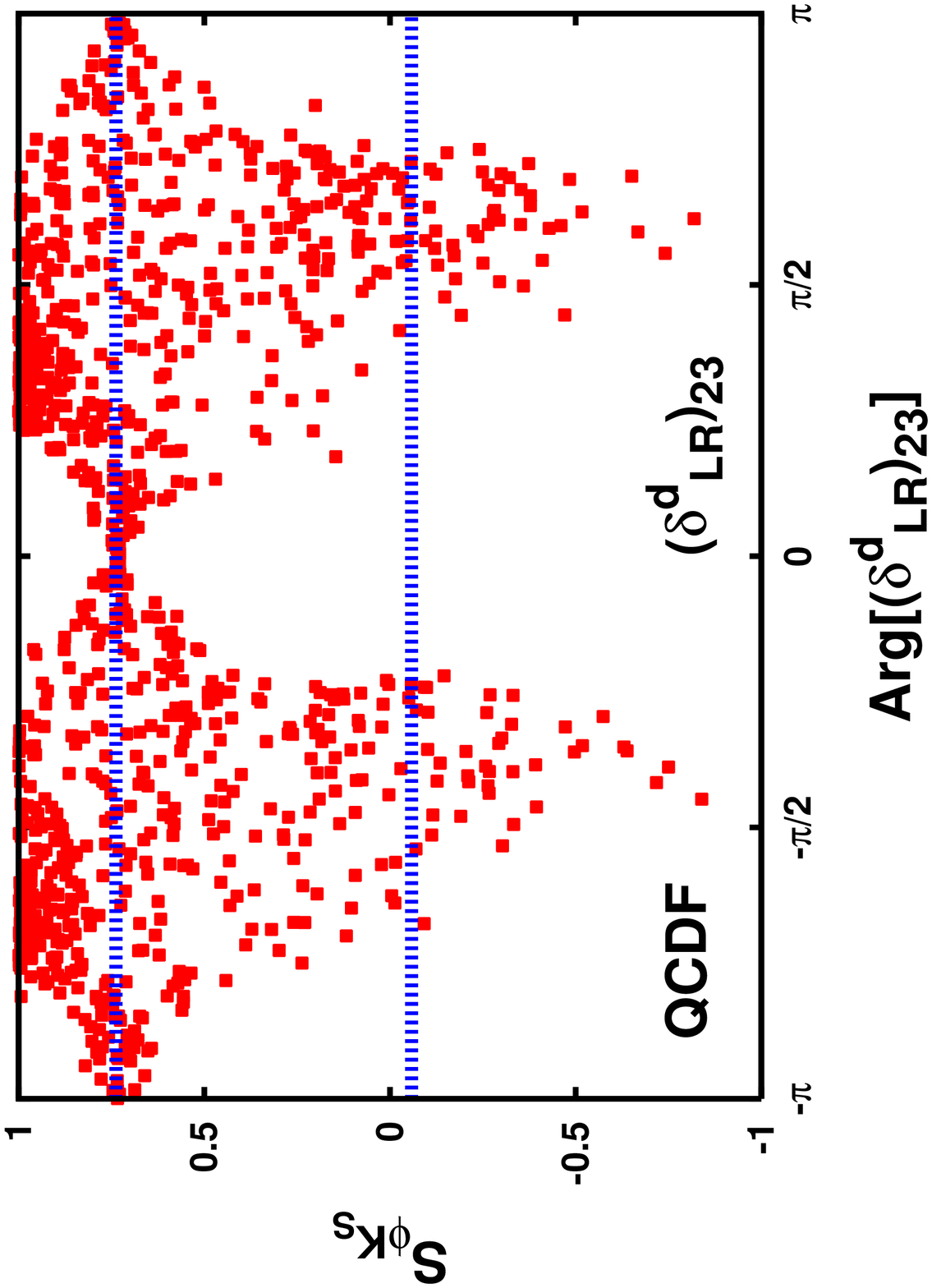}{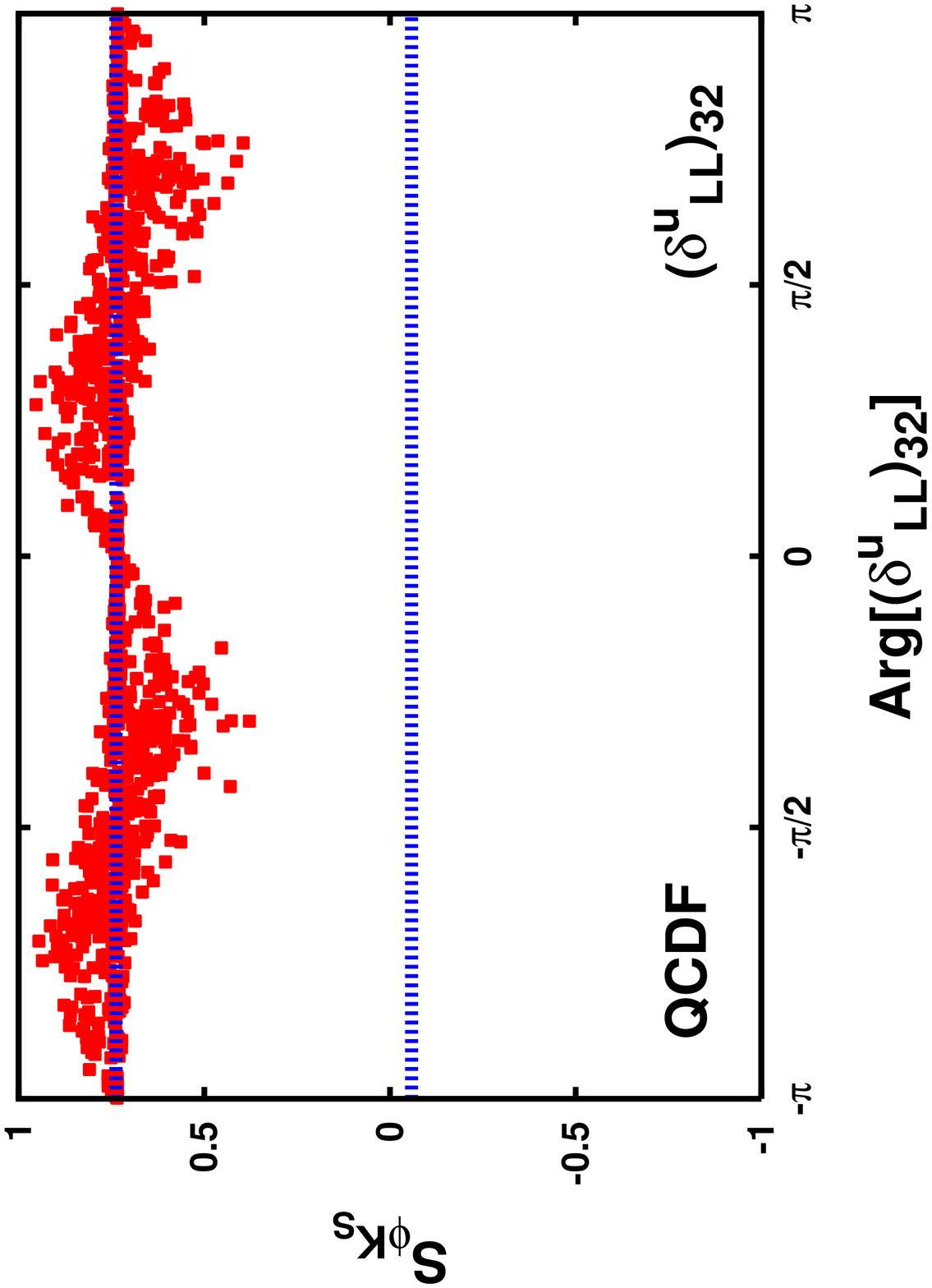}
\end{center}
\caption{\small $S_{\phi K_S}$ as a function of arg[$(\delta_{LR}^d)_{23}$]
(left) and  arg[$(\delta_{LL}^u)_{32}$] (right) with  gluino and chargino
contributions respectively. }
\label{Fig1}
\end{figure}

We present our numerical results for the gluino and chargino contributions to CP asymmetry
$S_{\Phi K_S}$  in Fig.~\ref{Fig1}. We plot the CP asymmetry as function of the
phase of $(\delta_{LR}^d)_{23}$ for gluino dominated scienario and $Arg[(\delta_{LL}^u)_{32}]$ for
the chargino dominanted model.
We have scanned over the relavant SUSY parameter space,
assuming SM central values as in table 1.
Namely, the average squark mass $\tilde{m}$, gluino
mass $m_{\tilde{g}}$.
Moreover we require that the SUSY spectra
satisfy the present experimental lower mass bounds.
In particular, $m_{\tilde{g}}> 200$ GeV,
$\tilde{m} > 300$ GeV. In addition, we scan over the real and
imaginary part of the corresponding mass insertions, by requiring that the
branching ratio (BR) of $b\to s \gamma$ and the $B^0-\bar{B}^0$ mixing constraints are satisfied
at  95\% C.L..
Also we have scanned over the
full range of the QCD factorization parameters $\rho_{A,H}$ and $\phi_{A,H}$,
We remind here that these parameters are taken into account for the (unknown) infrared
contributions in the hard scattering and annihilation diagrams respectively.

As can be seen from this figure, the gluino contributions proportional to  $(\delta_{LR}^d)_{23}$
have chances to drive $S_{\Phi K_S}$ toward the region of
larger and negative values. While in the  chargino
dominated scenario negative values of $S_{\phi}$ cannot be achieved.
The reason why extensive regions of negative values of $S_{\phi}$
are excluded here, is only due to the  $b\to s \gamma$ constraints.
Indeed, the inclusion of $(\delta_{LL}^u)_{32}$ mass insertion
can generate large and negative values of $S_{\phi}$,
by means of chargino contributions to chromomagnetic operator $Q_{8g}$
which are enhanced by terms of order $m_{\chi^{\pm}}/m_b$.
However, contrary to the gluino scenario,
chargino contributions to $C_{8g}$ are not enhanced by colour factors.
Therefore, large enhancements of the Wilson coefficient
$C_{8g}$, leave unavoidablly to the breaking of  $b\to s \gamma$ constraints.
As shown in Ref.\cite{emidio1},
by scanning over two mass insertion but requiring a common SUSY CP violating
phase, a sort of fine tuning to escape  $b\to s \gamma$ constraints is always possible,
and few points in the negative regions of $S_{\phi}$ can be approached.

\section{What happened to the $B \to \ep K_S$ process? }
Although $B \to \phi K_S$ and $B \to \eta^{\prime} K_S$ are very similar processes,
the parity of the final states can deviate the result. In  $B \to \phi K_S$, the contributions
from $C_i$ and $\tilde{C}_i$ to the decay amplitude are identically the same (with the same sign), while
in $B \to \ep K_S$, they have sign difference.
This can be simply understood by noticing that
\begin{equation}
\langle \phi K_S \vert Q_i \vert B \rangle =
\langle \phi K_S \vert \tilde{Q}_i \vert B \rangle\, .
\label{nfQQt}
\end{equation}
which is due to the invariance
of strong interactions under parity transformations, and
to the fact that initial and final states have same parity.
However, in case of $B \to \ep K_S$ transition, where the initial and final states have opposite parity,
we have
\begin{equation}
\langle \eta' K_S  \vert Q_i \vert B \rangle_{QCDF} = - \langle \eta' K_S
\vert \tilde{Q}_i
\vert B \rangle_{QCDF}.
\end{equation}

As a result, the sign of the $RR$ and $RL$ in the gluino contributions are different for
$B \to \phi K_S$ and $B \to \ep K_S$ \cite{KK2}. Using the same SUSY inputs
adopted in Eqs.~(\ref{Rch_Phi}), (\ref{Rgl_Phi}).
For gluino contributions we have
\bea
R^{QCDF}_{\eta^{\prime}}|_{\tilde{g}}\simeq
-0.07\,\times e^{i\,0.24}
 (\delta_{LL}^d)_{23}\,-\,
64(\delta_{LR}^d)_{23}+ 0.07\,\times e^{i\,0.24}
 (\delta_{RR}^d)_{23}\,+\,
64(\delta_{RL}^d)_{23}
\label{Rgl_eta}
\eea
while for chargino exchanges we obtain
\bea
R^{QCDF}_{\eta^{\prime}}|_{\chi^ {\pm}}&\simeq&
0.95\, (\delta_{LL}^u)_{32}\,-\,
0.025\times e^{-i\,0.19}
\, (\delta_{RL}^u)_{32}
\nonumber \\
&+&
0.21
\, (\delta_{LL}^u)_{31}\,-\,
0.006\times e^{-i\,0.19}
\, (\delta_{RL}^u)_{31}.
\label{Rch_eta}
\eea

\begin{figure}[tpb]
\begin{center}
\dofigs{3.1in}{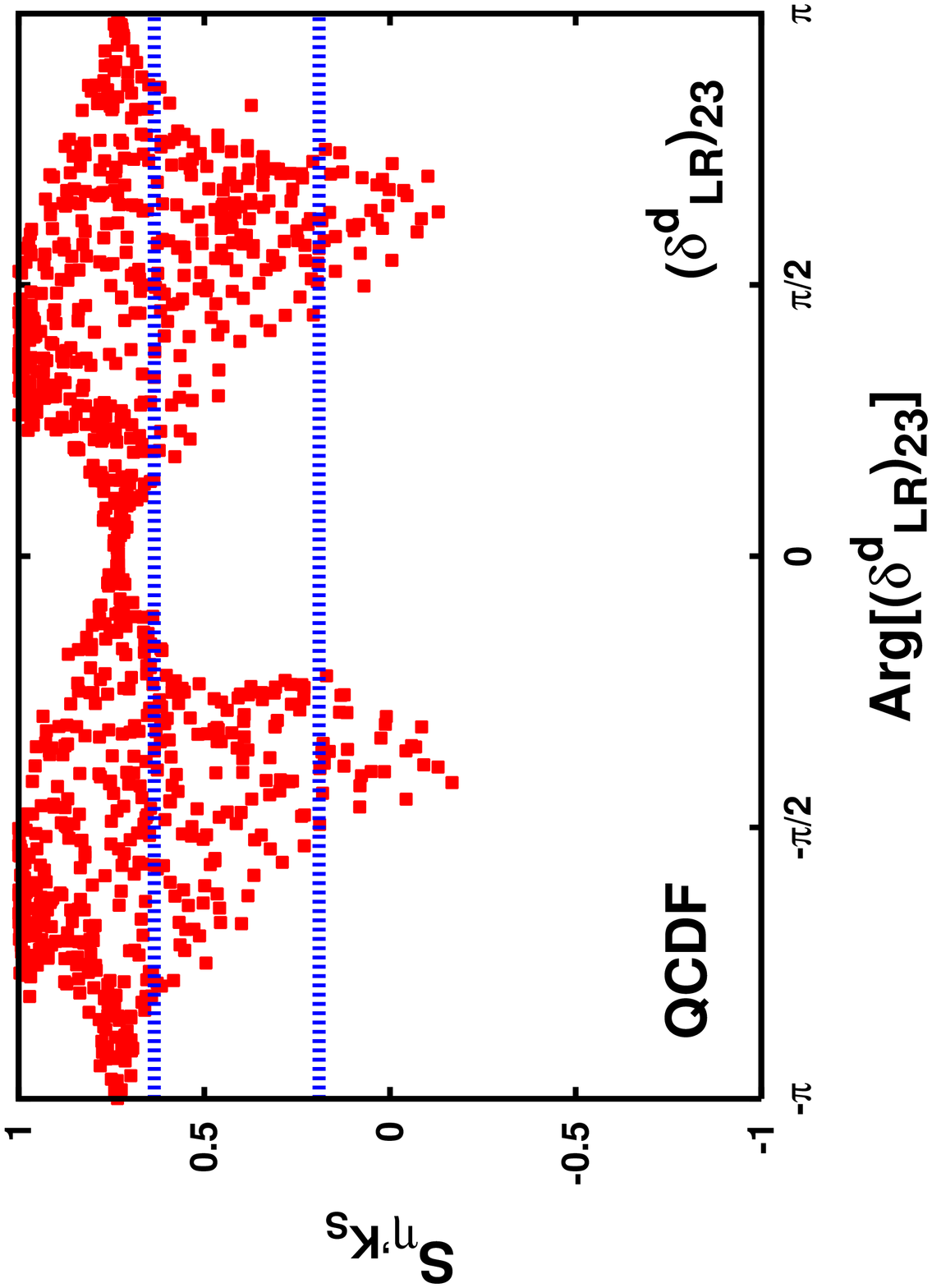}{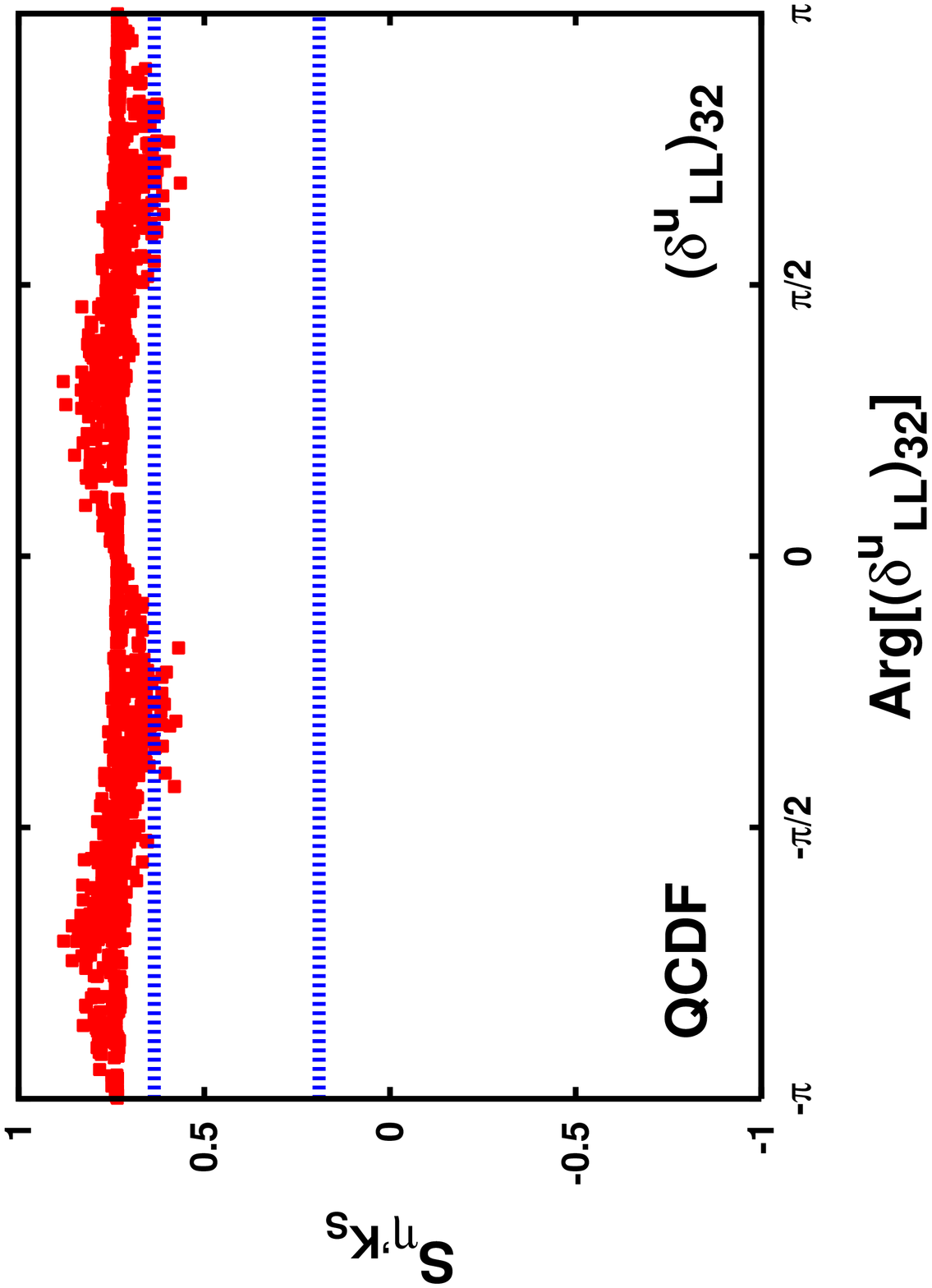}
\end{center}
\caption{\small $S_{\eta' K_S}$ as a function of arg[$(\delta_{LR}^d)_{23}$]
(left) and  arg[$(\delta_{LL}^u)_{32}$] (right) with  gluino and chargino
contributions respectively. }
\label{Fig2}
\end{figure}


We show our results for gluino and chargino contributions in Fig.~\ref{Fig2}, where we have just
extended the same analysis of $B\to \phi K_S$.
Same conventions as in Fig.~\ref{Fig1} for $B\to \phi K_S$ have been adopted here.
As we can see from these results, there is a depletion of the
gluino contribution in  $S_{\eta^{\prime}}$, precisely for the reasons
explained above. Negative regions are disfavoured, but a minimum of
$S_{\eta^{\prime}}\simeq 0$ can be achieved.
Respect the chargino contributions, it is clear that it can imply at most a
deviation from SM predictions of about $\pm 20$ \%.

\section{On the branching ratio of $B \to \ep K_S$}
In 1997, CLEO collaboration reported an unexpectedly large branching ratio \cite{CLEO}
\begin{equation}
Br^{\mbox{\tiny exp.}}(B^0\to K^0 \eta^{\prime}) =
(89^{+18}_{-16}\pm 9)\times 10^{-6}
\end{equation}
which is confirmed by Belle \cite{belle2} and BABAR \cite{babar2}:
\begin{eqnarray}
\mbox{BELLE}&=&(79^{+12}_{-16}\pm 8)\times 10^{-6}, \\
\mbox{BABAR}&=&(76.9 \pm 3.5\pm 4.4)\times 10^{-6}
\end{eqnarray}
Considering the theoretical prediction by the naive factorisation approximation
\begin{equation}
Br^{\mbox{\tiny theo.}}(B\to K \eta^{\prime}) \simeq 25\times 10^{-6},
\end{equation}
the experimental data is about factor of three large, thus, there have been
various efforts to explain this puzzle.
On one hand, new physics contributions have been discussed \cite{Kundu}. However, the enhancement
by new physics contributions through penguin diagrams ends up with large branching
ratios for all other penguin dominated processes. Therefore,  one needs a careful treatment
to enhance only $B \to \ep K$ process without changing the predictions for the other
processes.
On the other hand, since this kind of large branching ratio is observed only in
$B \to \ep K$ process, the gluonium contributors which only exist in this process
have been a very interesting candidate to solve the puzzle \cite{soni} \cite{emi}
though the amount of
gluonium in $\ep$ is not precisely known \cite{emi2}.  In this section, we discuss
the effect of our including SUSY contributions to the branching ratios for
$B \to \phi K$ and $B \to \ep K$.

Inclusion of the SUSY contributions modify the branching ratio as:
\[Br^{\mbox{\tiny SM + SUSY}}=Br^{\mbox{\tiny SM}}\times [1+2\cos\theta_{\mbox{\tiny SUSY}}R+R^2]
\]
where $R=|A^{\mathrm{SUSY}}|/|A^{\mathrm{SM}}|$.
As we have shown, to achieve a negative value of $S_{\phi K_S}$, we need
$\theta_{\mbox{\tiny SUSY}}\simeq -\pi/2$, which
suppresses the leading SUSY contribution.
On the other hand, the phase for $B \to \eta^{\prime} K $ is different from
the one for $\phi K_S$, as is discussed in the previous section.
In the following, we will analyze
the maximum effect one
can obtain from SUSY contribution to $BR(B\to \eta' K_S)$
with taking into account the
experimental limits on the $BR(B\to \phi K_S)$, $S_{\phi K_S}$
and $S_{\eta' K_S}$.

\begin{figure}[t]
\begin{center}
\dofigs{3.1in}{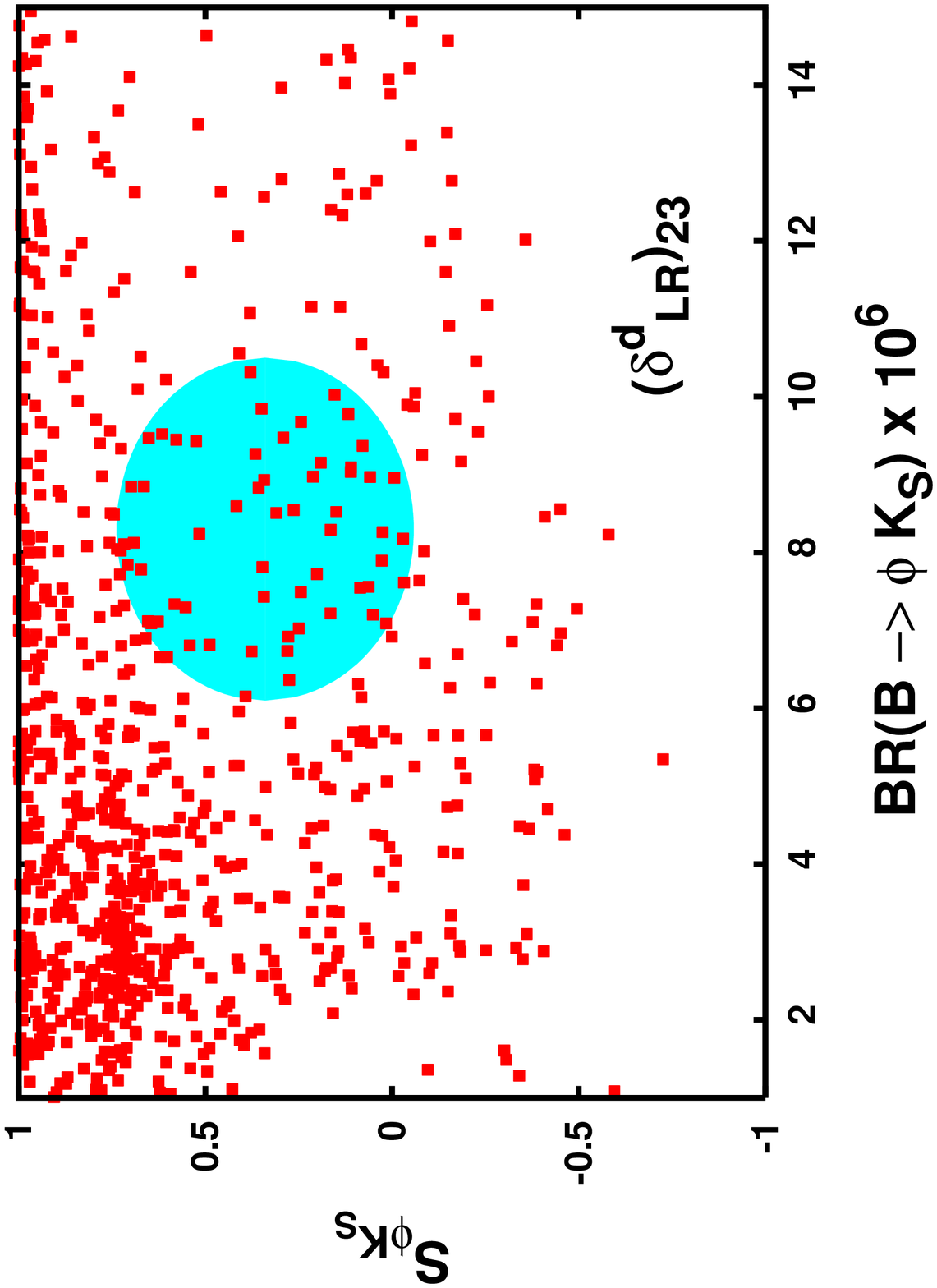}{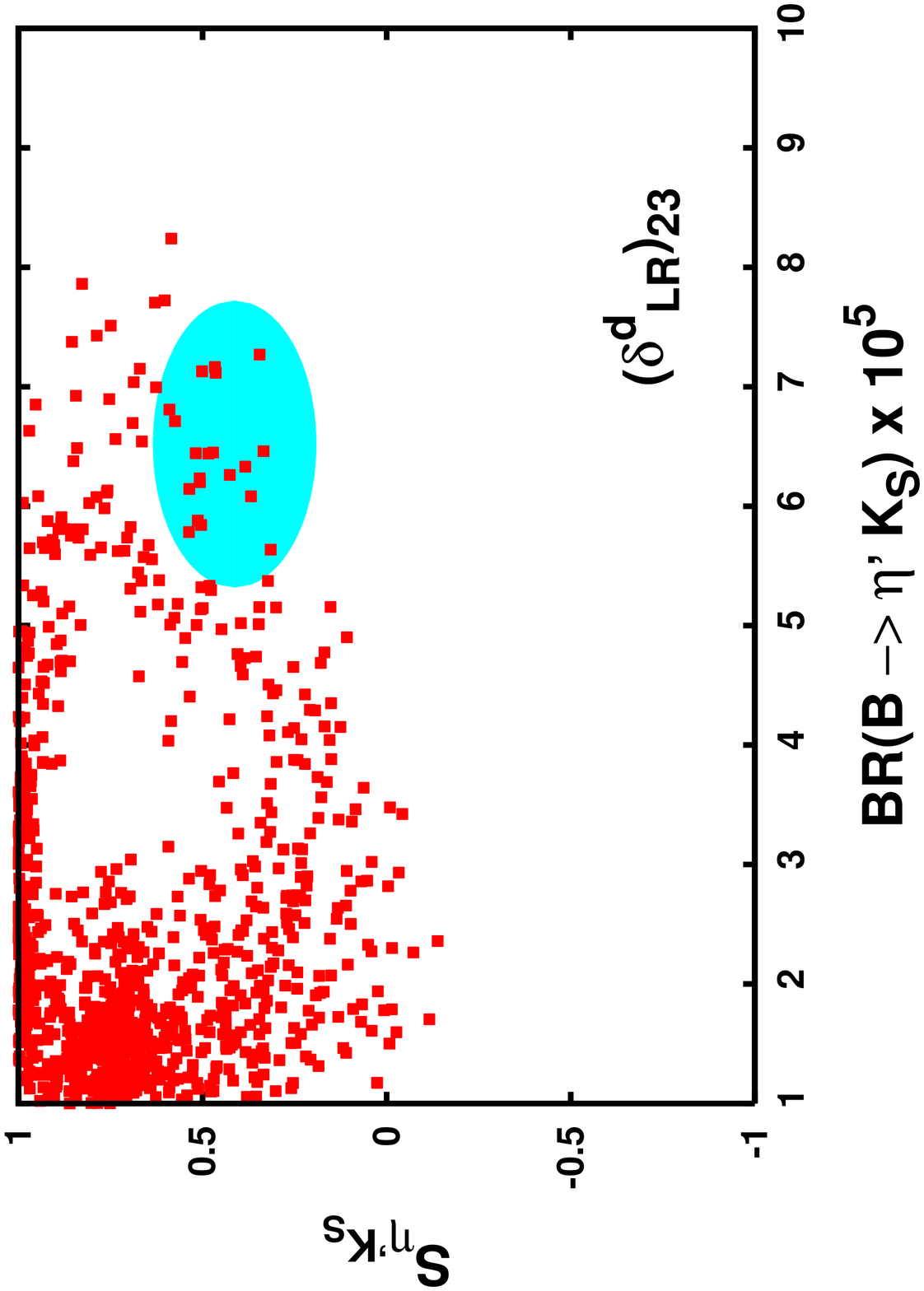}
\end{center}
\caption{\small
Correlations of $S_{\phi K_S}$ versus $BR(B\to \phi K_S )$ (left) and
$S_{\eta^{\prime} K_S}$ versus $BR(B\to \eta^{\prime} K_S )$
(right), for gluino contributions with one single
mass insertion $(\delta^d_{LR})_{23}$.
}
\label{BRS_DLR}
\end{figure}

In Fig.\ref{BRS_DLR} we plot the CP asymmetry versus the branching ratio for
$B\to \phi K_S$ and  $B\to \eta' K_S$. We consider the dominant
 gluino contribution
due to $(\delta^d_{LR})_{23}$ and scan over the other parameters as before.
One can see from this figure that, in the region of large negative
$S_{\phi}$  the $BR(B\to K_S \phi)$ is likely to be close to the SM prediction,
namely it is of  order $(2-5) \times  10^{-6}$. Larger values for the BR are also possible but
correspond to $S_{\phi K_S} \gsim -0.5$. In another word, if we consider the
central value of $BR(B\to K_S \phi)$ as $8\times 10^{-6}$, it is
predicted that  $S_{\phi K_S}$ likely to lie in the range $0 \lsim S_{K_S\phi} \lsim -0.5$.

Respect to the correlation between $S_{\eta' K_S}$ and $BR(B\to K_S \eta')$,
it is  remarkable that with just one mass insertion $S_{\eta' K_S}$ is likely
positive and  around $0.5$ which is quite compitable with the experimental results
and in this case with large $\mu \simeq m_b$, it is no longer needed to consider LR and RL mass
insertions semiltaneously to suppress $R_{\eta'}$ as explained in the previous section.
Furthermore, as can be seen from this figure, for $S_{\eta'K_S}\simeq 0.5$ the
$BR(B\to \eta' K_S)$ can be large as $80\times 10^{-6}$, {\it i.e}, it is enhanced by gluino
contribution  to more than 6 times the SM value and become compitable with the experimental result mentioned
above.

\begin{figure}[t]
\begin{center}
\dofigs{3.1in}{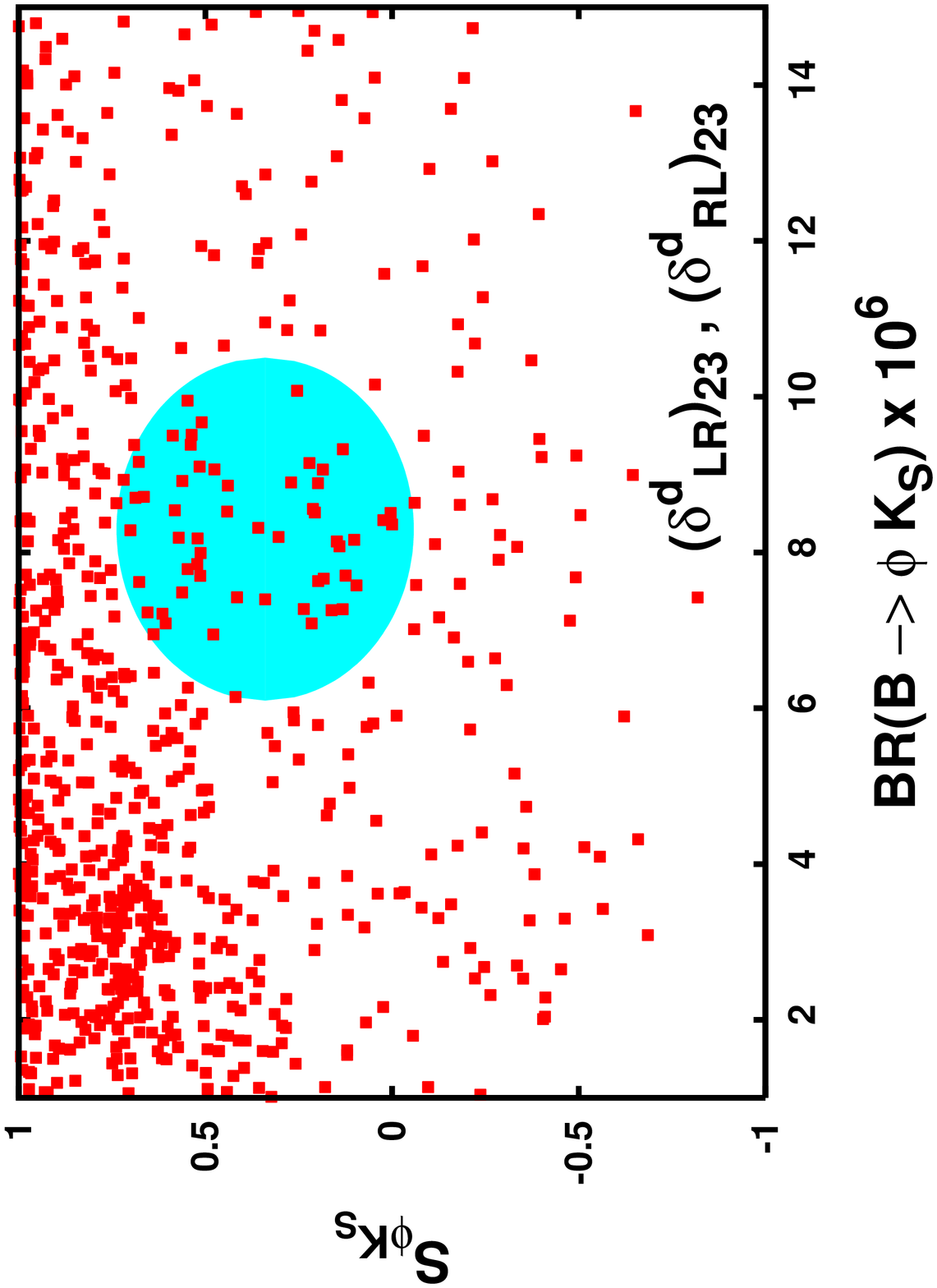}{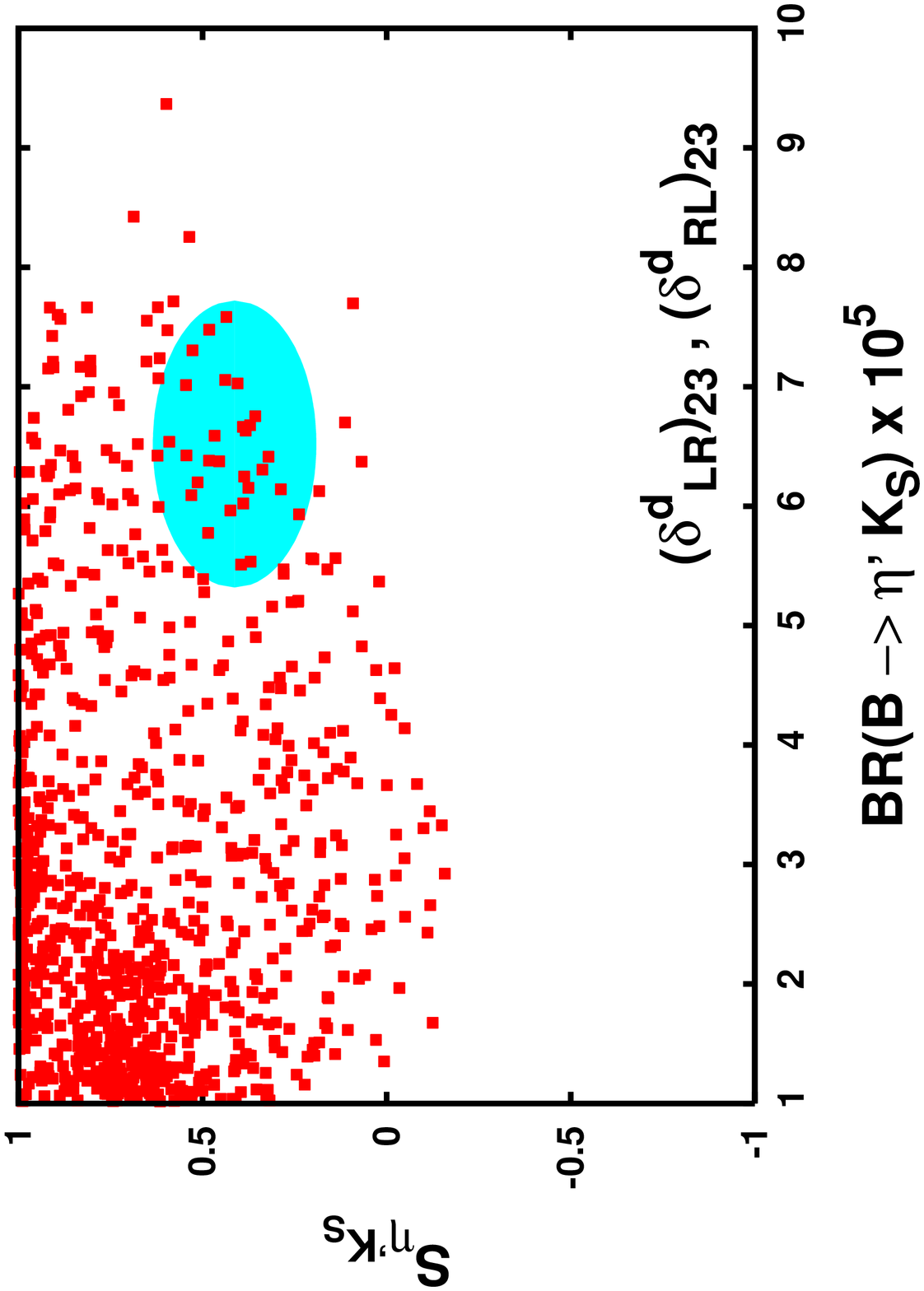}
\end{center}
\caption{\small As in Fig. \ref{BRS_DLR}, but
for gluino contributions with two
mass insertions $(\delta^d_{LR})_{23}$ and $(\delta^d_{RL})_{23}$.
}
\label{BRS_DLRRL}
\end{figure}

In Fig. \ref{BRS_DLRRL}, we present the correlation between  $S_{\phi K_S}$
and $BR(B\to \phi K_S)$ and also the correlation between
$S_{\eta' K_S}$ and $BR(B\to \eta' K_S)$.
Here we present the gluino contributions with two mass insertions
$(\delta^d_{LR})_{23}$ and
$(\delta^d_{LR})_{32}$. In this case, we can easily see that the gluino
contribution can saturate simultaneously both $S_{\phi K_S}$ and $BR(B\to \phi K_S)$ within
their experimental ranges. However, for $B\to \eta' K_S$, as expected the CP asymmetry becomes larger and
around the $\sin 2 \beta$ while its branching ratio is diminished. Now it is of order $(20-40)
\times 10^{-6}$ which is smaller than the experimental measurments.

\begin{figure}[t]
\begin{center}
\dofigs{3.1in}{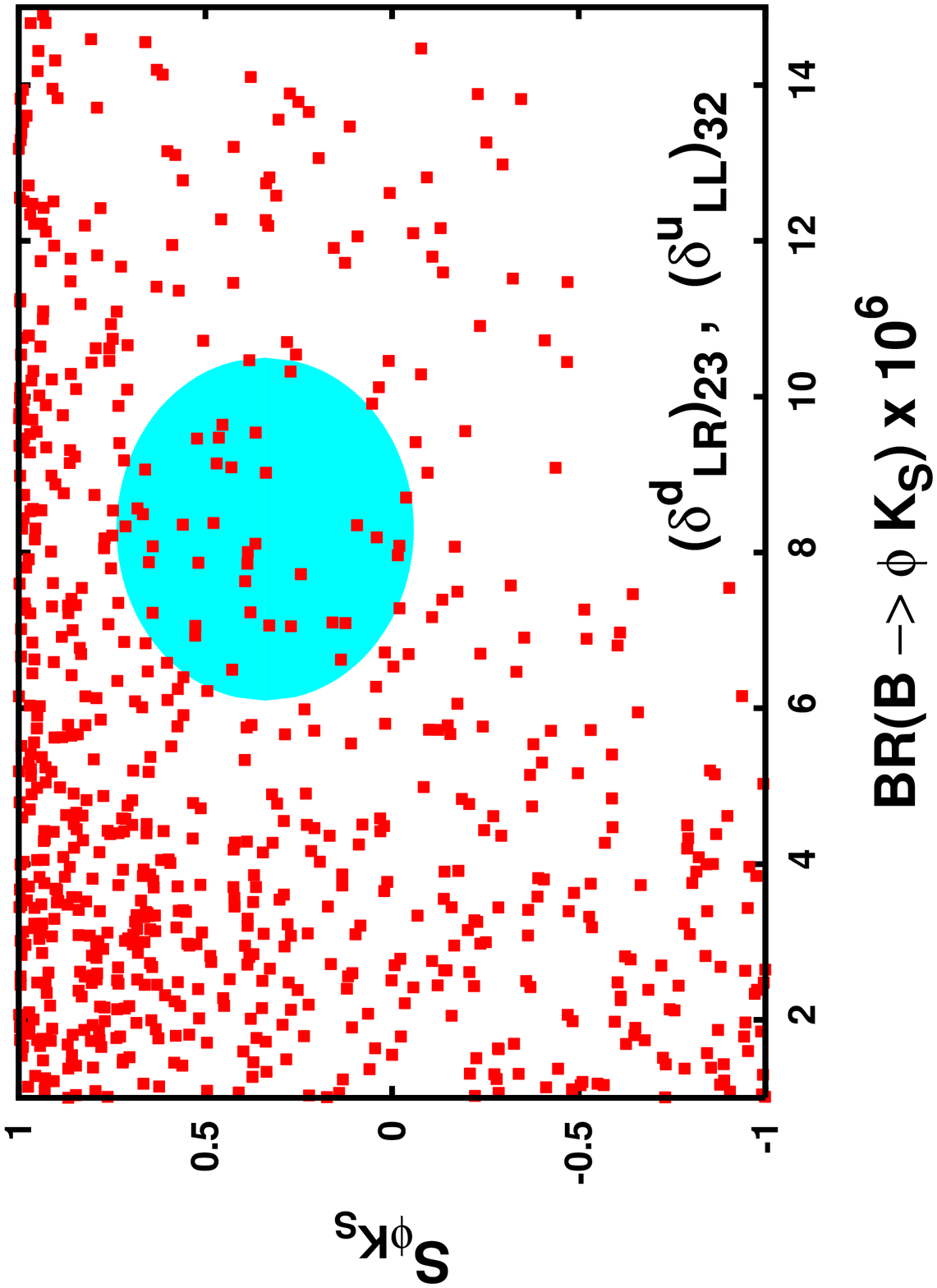}{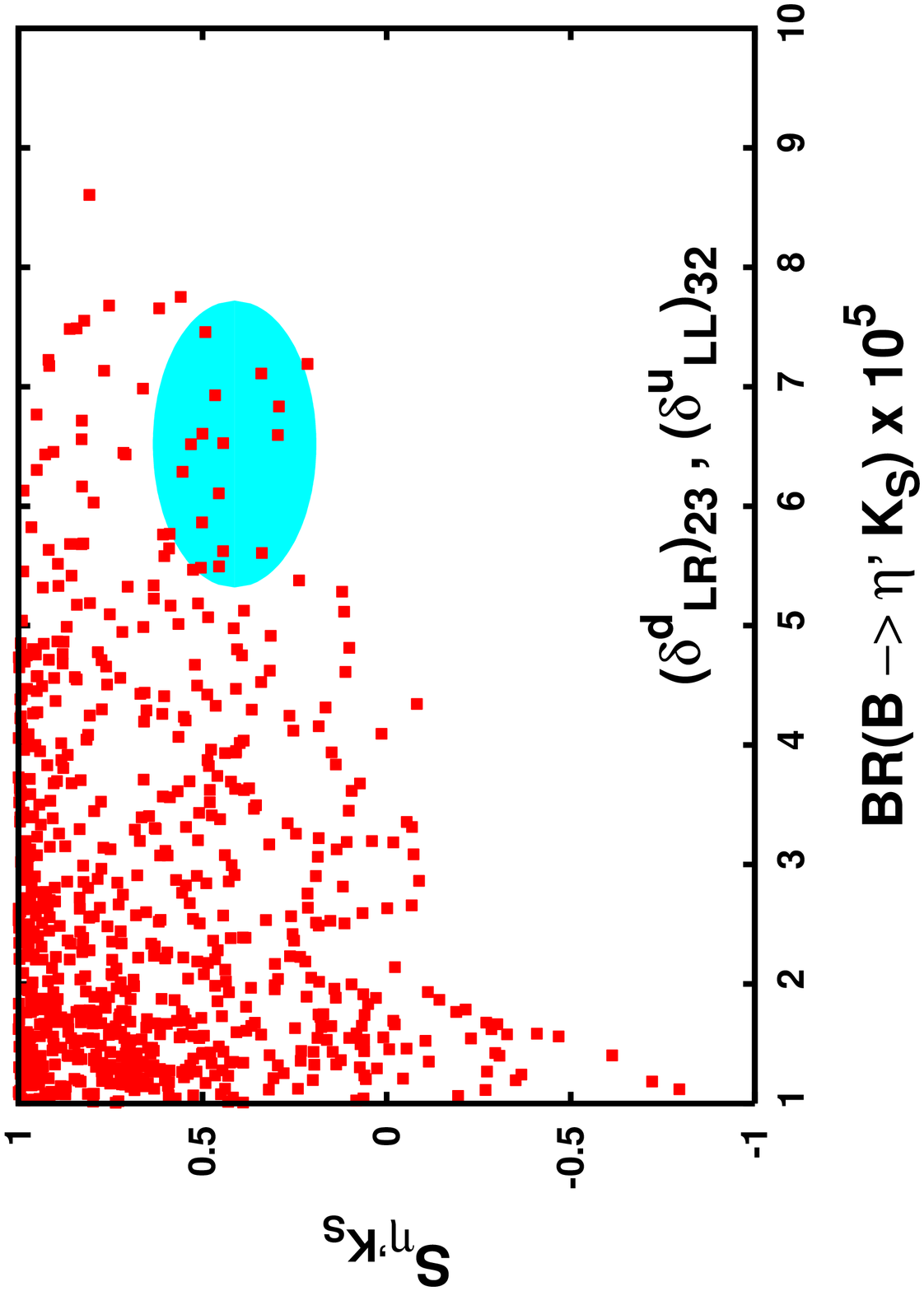}
\end{center}
\caption{\small As in Fig. \ref{BRS_DLR}, but
for gluino and chargino contributions with
mass insertions $(\delta^d_{LR})_{23}$ and $(\delta^u_{LL})_{32}$ respectively.
}
\label{BRS_DLR_ULL}
\end{figure}

The combination effects from gluino and chargino on $BR(B\to \phi K_S)$ and
$BR(B\to \eta' K_S)$ are shown in Fig.\ref{BRS_DLR_ULL}. We present the
CP asymmetry versus the branching ratio for each process. We scan on
the allowed range of the most relevant mass insertions for these two contributions:
$(\delta^d_{LR})_{23}$ and $(\delta^u_{LL})_{32}$. We also vary the other parameters as in
the previous figures. The message of this figure is that with both gluino and chargino
we can easily accommodate the experimental results for the CP asymmetries and the branching
ratios of $BR(B\to \phi K_S)$ and $BR(B\to \eta' K_S)$. It is important to stress that
the striengent bound on $(\delta^u_{LL})_{32}$ from the experimental limits on $BR(B\to X_s \gamma)$
are relaxed when one consider both gluino and chargino contributions, which comes with different sign.
Now some configuration with large $\tan\beta$ are allowed and therefore chragino can contribute
significantly to the CP asymmetries $S_{\phi K_S}$ and $S_{\eta' K_S}$.
It is also remarkable that in this scienario, the value of the branching ratio
$BR(B\to \eta^{\prime} K_S )$ can be of order $60\times 10^{-6}$ which is compitable with
the central value of the experimental results.

\section{Conclusions}
We studied the supersymmetric contributions to the CP asymmetry of
$B \to \phi K_S$ and $B \to \ep K_S$ in a model independent way.
We found that the observed large discrepancy between $S_{J/\psi K_S}$ and
$S_{\phi K_S}$ can be explained within some SUSY models with large $(\delta_{LR})_{23}$
or $(\delta_{RL})_{23}$ mass insertions.
We showed that the SUSY contributions of $(\delta_{RR})_{23}$ and
$(\delta_{RL})_{23}$ to  $B \to \phi K_S$ and $B \to \eta^{\prime} K_S$
have different signs. Therefore, the current observation,
$S_{\phi K_S} < S_{\ep K_S}$, favours the $(\delta_{RR,RL})_{23}$ dominated models.
We also discussed the SUSY contributions to the branching ratios.
We showed that negative $S_{\phi K_S}$ and small SUSY effect to $Br(B \to \phi K)$
can be simultaneously achieved. On the other hand, we showed that
SUSY contribution itself may {\it not} solve the puzzle of the large branching ratio
of $B \to \eta^{\prime} K_S$.



\begin{thebibliography}{99}
%
\bibitem{Abel:2001vy}
S.~Abel, S.~Khalil and O.~Lebedev,
Nucl.\ Phys.\ B {\bf 606}, 151 (2001).
\bibitem{Bfact} B.~Aubert {\it et al.} [BaBar Collaboration],
Nucl. Instrum. Meth. {\bf A 479} (2002) 1; S.~Mori {\it et al.}
[Belle Collaboration], Nucl.\ Instrum.\  Meth. {\bf A 479} (2002)
117.

\bibitem{PhiKs_NP}
R.~Fleischer and T.~Mannel, Phys.\ Lett. {\bf B 511} (2001) 240;
C.~W.~Chiang and J.~L.~Rosner,  Phys.\ Rev.\ {\bf D 68} (2003)
014007;

\bibitem{KM}
S.~Khalil and R.~Mohapatra, Nucl.\ Phys.\ {\bf B 695} (2004) 313

\bibitem{PhiKs_Rparity}
A.~Datta,  Phys.\ Rev.\ {\bf D 66} (2002) 071702; B.~Dutta,
C.~S.~Kim and S.~Oh,  Phys.\ Rev.\ Lett.\ {\bf 90} (2003) 011801;
A.~Kundu and T.~Mitra,  Phys.\ Rev.\ {\bf D 67} (2003) 116005;
B.~Dutta, C.S.~Kim, S.~Oh, G.~Zhu, hep-ph/0312388, hep-ph/0312389.

\bibitem{PhiKs_gluino}
E.~Lunghi and D.~Wyler, Phys.\ Lett. {\bf B 521} (2001) 320;
M.~B.~Causse, arXiv:hep-ph/0207070; G.~Hiller, Phys.\ Rev.\ {\bf D
66} (2002) 071502; M.~Ciuchini and L.~Silvestrini,  Phys.\ Rev.\
Lett.\ {\bf 89} (2002) 231802; S.~Khalil and E.~Kou,  Phys.\ Rev.\
{\bf D 67} (2003) 055009; K.~Agashe and C.~D.~Carone, Phys.\ Rev.\
{\bf D 68} (2003) 035017; G.~L.~Kane, P.~Ko, H.~b.~Wang, C.~Kolda,
J.~h.~Park and L.~T.~Wang;  Phys.\ Rev.\ Lett.\ {\bf 90} (2003)
141803; C. Dariescu, M.A. Dariescu, N.G. Deshpande, D.K. Ghosh,
Phys.\ Rev.\ {\bf D 69} (2004) 112003; M.~Ciuchini, E.~Franco,
G.~Martinelli, A.~Masiero, M.~Pierini, L.~Silvestrini,
hep-ph/0407073; J.~F.~Cheng, C.~S.~Huang and X.~H.~Wu,
Nucl.\ Phys.\ B {\bf 701}, 54 (2004).

\bibitem{PhiKs_sugra}
Z.~Xiao and W.~Zou, hep-ph/0407205.


\bibitem{CFMS}
M.~Ciuchini, E.~Franco, A.~Masiero, L.~Silvestrini,  Phys.\ Rev.\
{\bf D 67} (2003) 075016, Erratum Phys.\ Rev.\ {\bf D 68} (2003)
079901.

\bibitem{KK} S.~Khalil and E.~Kou, Phys.\ Rev.\ Lett. {\bf 91} (2003)
241602;\\
http://www.slac.stanford.edu/gen/meeting/ssi/2002/kagan1.html.

\bibitem{chargino} S.~Baek,  Phys.\ Rev.\ D {\bf 67} (2003) 096004;
Y.~Wang, Phys.\ Rev.\ {\bf D 69} (2004) 054001.

\bibitem{emidio1}
D.~Chakraverty, E.~Gabrielli, K.~Huitu and S.~Khalil,
Phys.\ Rev.\ D {\bf 68}, 095004 (2003) [arXiv:hep-ph/0306076].

\bibitem{emidio2}
E.~Gabrielli, K.~Huitu and S.~Khalil, hep-ph/0407291.

\bibitem{hpenguin} J.F.~Cheng, C.S.~Huang, X.H.~Wu, Phys. Lett.
{\bf B 585} (2004) 287.

\bibitem{AB} A.J.~Buras, hep-ph/0210291; A.J.~Buras, F. Parodi, and A. Stocchi,
JHEP {\bf 0301} (2003) 029.

\bibitem{giorgi}
M. A.Giorgi (BaBar collaboration), plenary talk at XXXII Int.
Conference on High Energy Physics, Beijing, China, August 16-22,
2004, http://ichep04.ihep.ac.cn/

\bibitem{sakai}
Y. Sakai (Belle collaboration), plenary talk at XXXII Int.
Conference on High Energy Physics, Beijing, China, August 16-22,
2004, http://ichep04.ihep.ac.cn/

\bibitem{phi_babar} B. Aubert et al., BaBar Collaboration, hep-ex/0403026.

\bibitem{phieta_belle}
K.~Abe {\it et al.}, Belle Collaboration, Phys.\ Rev.\ Lett.\ {\bf
91} (2003) 261602.

\bibitem{hfag} Heavy Flavor Averaging Group,
http://www.slac.stanford.edu/xorg/hfag/.

\bibitem{etaBa} B.~Aubert {\it et al.},
BaBar Collaboration, Phys.\ Rev.\ Lett.\ {\bf 91} (2003) 161801.

\bibitem{GNR}
Y. Grossman, Y. Nir, and R. Rattazzi, Adv. Serv. Direct. High Energy
Phys. {\bf 15} (1998) 755; hep-ph/9701231.

\bibitem{HallRaby} 
L.~J.~Hall, V.~A.~Kostelecky and S.~Raby,
Nucl.\ Phys.\ B {\bf 267} (1986) 415.


\bibitem{KK2} 
S.~Khalil and E.~Kou,
arXiv:hep-ph/0303214.


\bibitem{CLEO} 
S.~J.~Richichi {\it et al.}  [CLEO Collaboration],
arXiv:hep-ex/9908019.

\bibitem{belle2} 
K.~Abe {\it et al.}  [Belle Collaboration],
Phys.\ Lett.\ B {\bf 517}, 309 (2001)
[arXiv:hep-ex/0108010].

\bibitem{babar2}
B. Aubert et al., BaBar Collaboration,  Phys.\ Rev.\ D {\bf 69}
(2004) 0111102; B. Aubert et al., Babar Collaboration , Phys.\
Rev.\ Lett.\ {\bf 91} (2003)  161801.


\bibitem{Kundu} 
A.~Kundu,
Pramana {\bf 60} (2003) 345
[arXiv:hep-ph/0205100].

\bibitem{soni}  
D.~Atwood and A.~Soni,
Phys.\ Lett.\ B {\bf 405} (1997) 150
[arXiv:hep-ph/9704357].

\bibitem{emi} 
M.~R.~Ahmady, E.~Kou and A.~Sugamoto,
Phys.\ Rev.\ D {\bf 58} (1998) 014015
[arXiv:hep-ph/9710509].

\bibitem{emi2} 
E.~Kou,
Phys.\ Rev.\ D {\bf 63} (2001) 054027
[arXiv:hep-ph/9908214].



\end{thebibliography}
\end{document}